\documentclass[twocolumn,amsmath,amssymb]{snp}
\usepackage{graphicx}
\usepackage{dcolumn}
\usepackage{bm}
\newcommand{\nn}{\nonumber}
\newcommand{\be}{\begin{equation}}
\newcommand{\ee}{\end{equation}}
\newcommand{\bea}{\begin{eqnarray}}
\newcommand{\eea}{\end{eqnarray}}

\newcommand{\om}{\omega}    
\newcommand{\vk}{\vec k}

\begin{document}

\title{{\Large  Role of interaction on electrical conductivity of QGP in presence of magnetic field}}

\author{\large Ankita Mishra$^{1,*}$, Souvik Paul$^{2}$,  Jayanta Dey$^3$, Sarthak Satapathy$^3$,
Sabyasachi Ghosh$^3$}
\email{m02ankita@gmail.com}
\affiliation{$^1$Department of Mechanical Engineering, Guru Ghasidas University, Bilaspur 495009, India}
\affiliation{$^2$Department of Physical Sciences,Indian Institute of Science Education 
and Research Kolkata, Mohanpur, West Bengal 741246, India}
\affiliation{$^3$Indian Institute of Technology Bhilai, GEC Campus, Sejbahar, Raipur 492015, 
Chhattisgarh, India}
\maketitle

Recently, through lattice quantum chromo dynamics (LQCD), Bali et al.~\cite{Bali2} 
have addressed thermodynamics of quark gluon plasma (QGP)
in presence of magnetic field, where thermodynamical quantities like pressure, energy
density, entropy density get suppresing values with respect to their Stephan-Boltzmann (SB) limit.
The suppression increases as temperature ($T$) and magnetic field ($B$) reduce. Ref.~\cite{LQCD_QGP} can realize
the reduction of thermodynamical quantities at $B=0$ as reduction of degeneracy factors, when we move 
from high $T$ QGP phase to low $T$ hadronic matter (HM) phase. The present work has extended that 
interacting QGP picture for finite magnetic field case by finding a gross temperature and 
magnetic field dependent degeneracy factor $g(T,B)$.
 \begin{figure}
 \centering
 \includegraphics[scale=0.25]{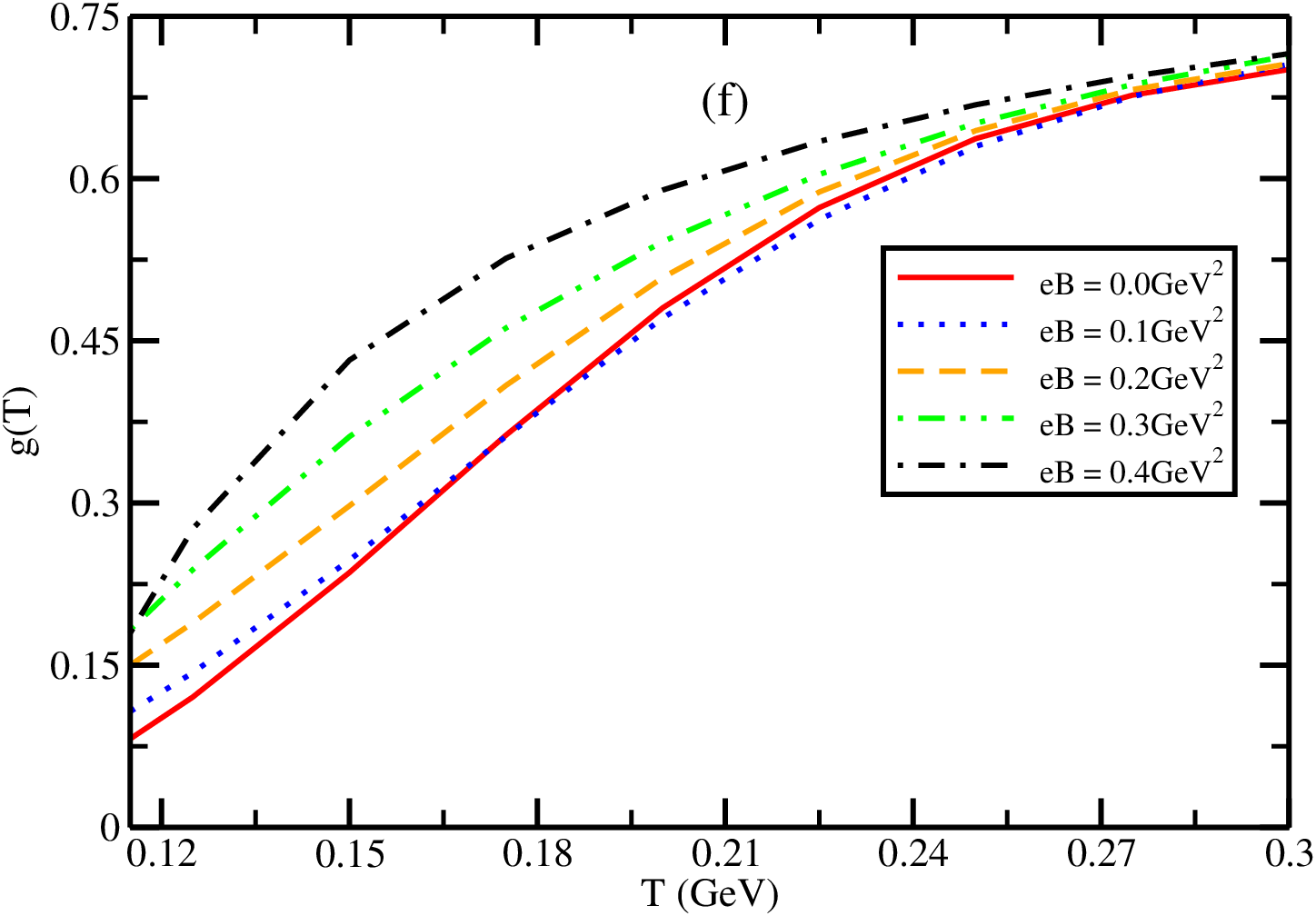} 
 \caption{Degeneracy factors $g(T)$ at for different strengths of magnetic field are obtained
 by matching LQCD data~\cite{Bali2}.}
 \label{gvsT}
 \end{figure}

By matching the LQCD data points of thermodynamical quantities of QGP 
for $eB=0$, $0.1$, $0.2$, $0.3$, $0.4$ GeV$^2$~\cite{Bali2}, we find a parametric form
\begin{equation}
g(T)= a_0 - \frac{a_1}{e^{a_2(T-0.17)}+ a_3}~,
\label{g_a0123}
\end{equation}
where values of $a_{0,1,2,3}$ for different $eB$'s are given in Table.~(I).
%
\begin{table}
	\begin{center}
	\caption{Different values of $a_{0,1,2,3}$, given in Eq.~(\ref{g_a0123}) for different magnetic
	field strengths}
	\label{tab:table1}
	\begin{tabular}{ |c|c|c|c|c| } 
		\hline
		$\mathbf{eB(GeV^2)}$ & $\mathbf{a_0}$ & $\mathbf{a_1}$ & $\mathbf{a_2}$ & $\mathbf{a_3}$\\ 
		 \hline
		0.0 & 0.834334 & 0.122845 & 3.87082 & -0.620074\\ 
		0.1 & 0.751006 & 0.797544 & 21.6385 & 0.934298\\ 
		0.2 & 0.743269 & 0.615513 & 21.2978 & 0.727439\\
		0.3 & 0.786970 & 0.404290 & 12.8704 & 0.17983\\
		0.4 & 0.864099 & 0.0778032 & 2.03964 & -0.77928\\
		\hline
	\end{tabular}
\end{center}
\label{tab1}
\end{table}
These $g(T)$ for different $eB$'s are plotted in Fig.~(\ref{gvsT}), where we
notice that 
by reducing the degeneracy factor of QGP with reducing the temperature and magnetic
field, one can properly map quark-hadron phase transition along temperature and magnetic
field axis as described in Fig.~(\ref{gvsT})
%
\begin{figure}
 \centering
 \includegraphics[scale=0.25]{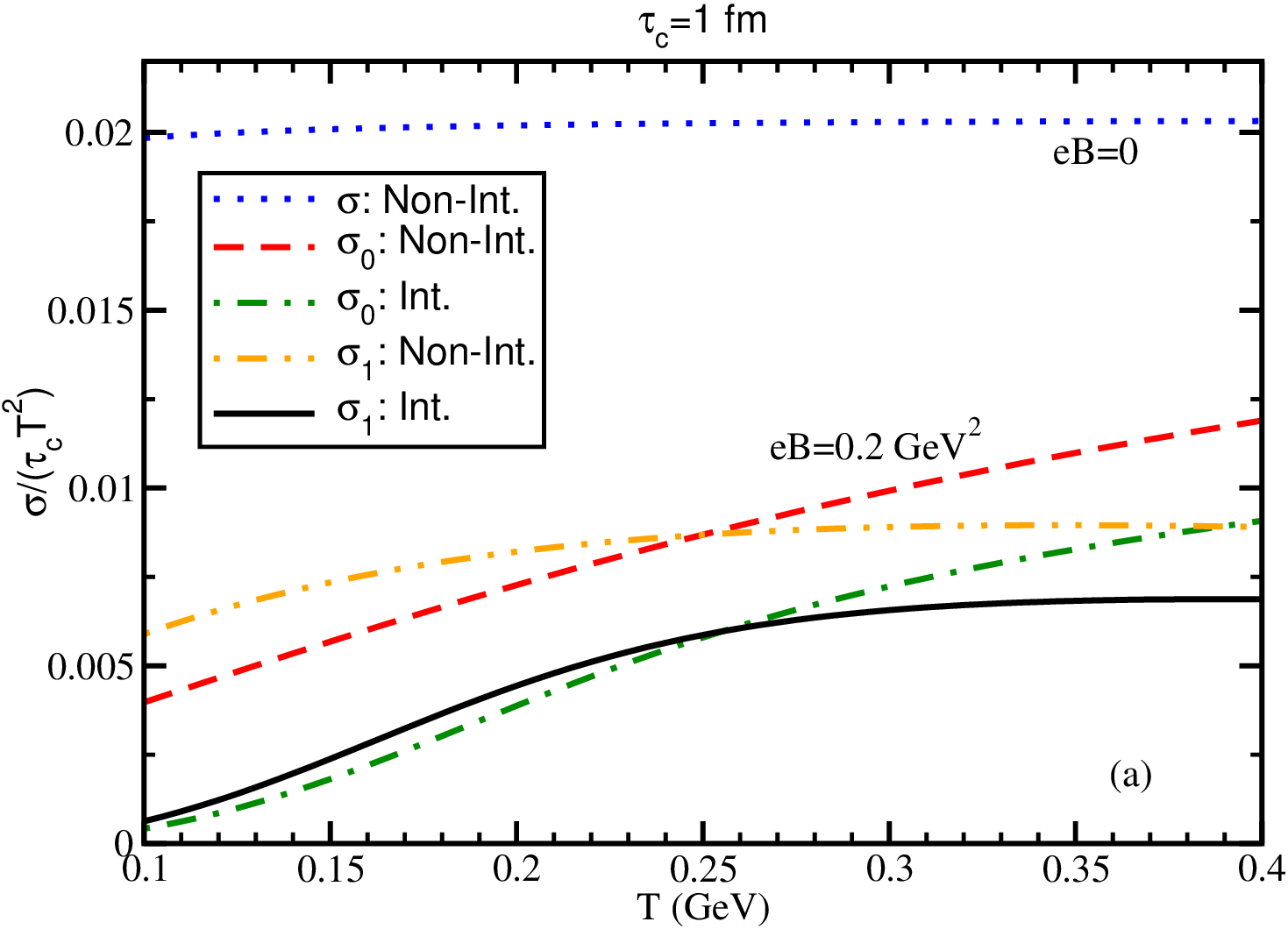}
 \includegraphics[scale=0.25]{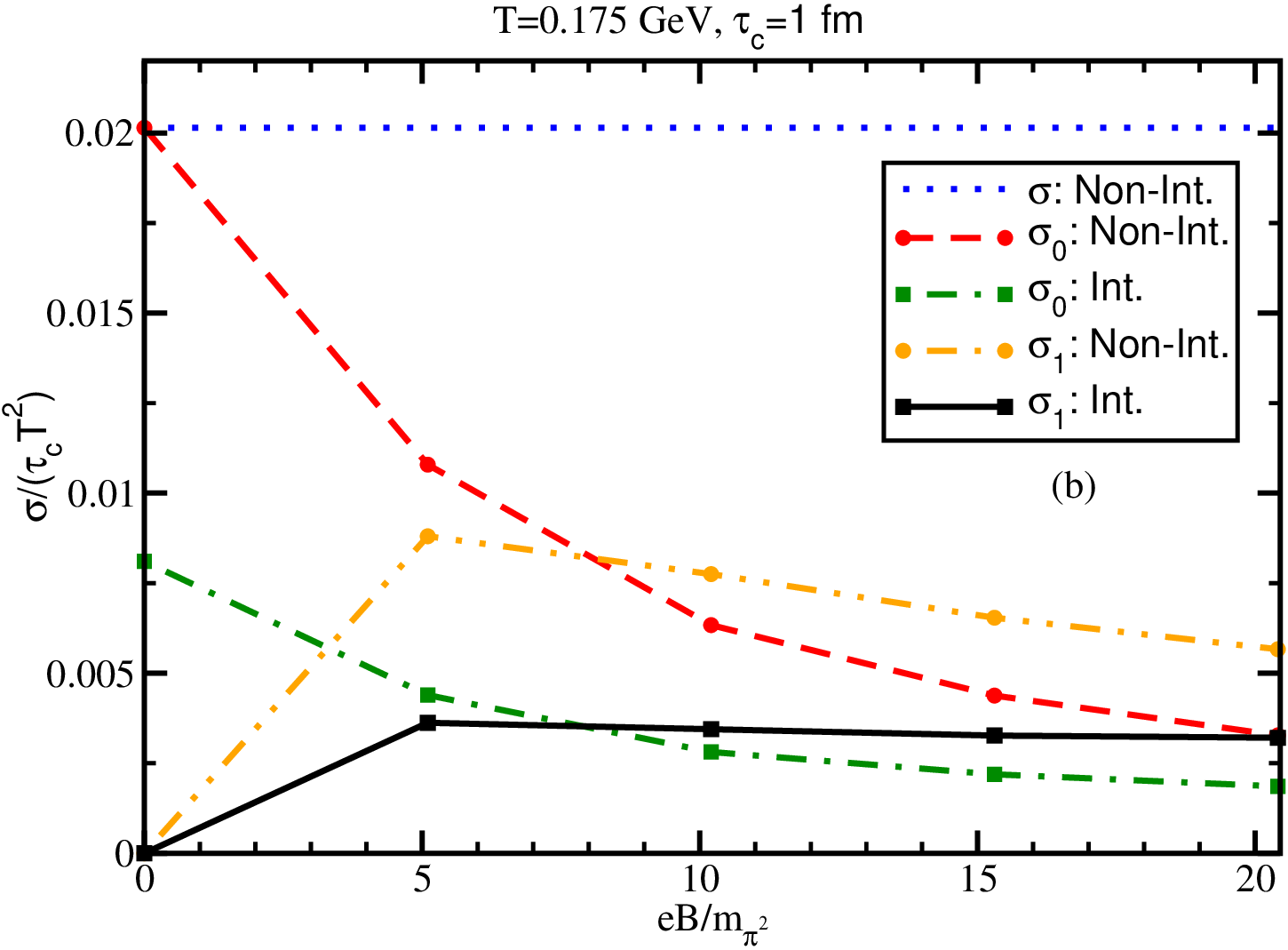} 
 \caption{Temperature and magnetic field dependence of electrical conductivity
 of non-interacting and interacting QGP system.}
 \label{sig_TB}
 \end{figure}

After building the quasi-particle description of interacting QGP system in presence
of magnetic field, next we are interested to find their impact on its transport Coefficient
like electrical conductivity, which play a vital role in time evolotion of magnetic
field $B(t)$~\cite{Tuchin:2013ie}.

In absence of magnetic field, medium follow isotropic transport properties,
for which we will get single component of electrical
conductivity ($\sigma$), but they become multi-component in presence of magnetic field.
We will get two main electrical conductivity components - normal ($\sigma_{0}$) 
and Hall ($\sigma_{1}$) coefficients, whose standard relativistic expressions are given below
\bea
\sigma_{0,1} &=&\sum_{i=u,d,s}\frac{g{\tilde e}_i^2\beta}{3}\int \frac{d^3k}{(2\pi)^3}\frac{\vk^2}{(\vk^2+m_i^2)}\tau_c
\nn\\
&&~~~\Big(\frac{(\tau_c/\tau_B)^{0,1}}{1+(\tau_c/\tau_B)^2}\Big)f_i(1- f_i)
\label{el_B_QGP}
\eea
where $g=12$, ${\tilde e}_{u,d,s}^2=\frac{4e^2}{9}$, $\frac{e^2}{9}$, $\frac{e^2}{9}$,
$m_{u,d,s}=0.005$, $0.005$, $0.100$ GeV, $f_i$ is Fermi-Dirac distribution function.
In above equation, $\tau_B=\omega/(eB)$ is appeared as a new time scale due to magnetic field
along with the relaxation time $\tau_c$, already existed at $B=0$. 
One should notice that the charge neutral gluon does not have any role in electrical conductivity.

Using Eq.~(\ref{el_B_QGP}) at $B=0$, $\sigma/(\tau_c T^2)$ for non-interacting QGP system is obtained and
shown by dotted line in Fig.~(\ref{sig_TB}), which looks like straight horizontal line of SB 
limit for thermodynamical quantities. Next we have plotted normal (dash line) and 
Hall (dash-double-dotted line) components of electrical conductivity $\sigma_{0,1}$ 
for non-interacting QGP, which exhibit suppresed value with respect to $\sigma$. 
When we plug in the interaction through $g(T, B)$ in Eq.~(\ref{el_B_QGP}), conductivity
will be suppresed more. From Fig.\ref{sig_TB}(b), one can see that
at $B\rightarrow 0$, normal components $\sigma_0$ of transport coefficients 
coincide with their without field isotropic value $\sigma$ and
Hall components $\sigma_1$ becomes zero as expected. 
As $B$ becomes non-zero and increases Hall components
also become non-zero and grow up, whereas normal components are reduced. The $B$ dependence
is hidden in $\tau_B=\om/({\tilde e}B)$ and approximately $\sigma_0\propto\frac{1}{1+(\tau_c/\tau_B)^2}$
and $\sigma_1\propto\frac{(\tau_c/\tau_B)}{1+(\tau_c/\tau_B)^2}$ functional dependence
are reflected in in Figs.~\ref{sig_TB}(b). Analyzing the Hall component
anisotropic factor $\frac{(\tau_c/\tau_B)}{1+(\tau_c/\tau_B)^2}$, we can get peak its value
around $\tau_B\approx\tau_c$ and then it can be reduced with $B$. 
A detail investigation can be seen in Ref.~\cite{QGP_B}. The message of increasing
electrical conductivity (normal component) with decreasing magnetic field might prevent decay of external
magnetic field in RHIC or LHC experiments. according to Ref.~\cite{Tuchin:2013ie}. 
Although a details phenomenological studies is required
to check this by using our calculated $\sigma^0(T, B)$.

\end{document}